\documentclass[11pt]{article}
\def\trace{\mbox{tr}\;}

\def\boldpi{\mbox{\boldmath$\pi$}}
\def\boldxi{\mbox{\boldmath$\xi$}}

\def\longminus{-\!\!\!-\!\!\!-\!\!\!-\!\!\!-}

\begin{document}

\begin{titlepage}
\begin{flushright}
CALT-68-2477\\
MPG-10/04
\end{flushright}

\begin{center}
{\Large\bf $ $ \\ $ $ \\
Slow evolution of nearly-degenerate \\ 
extremal surfaces}\\
\bigskip\bigskip\bigskip
{\large Andrei Mikhailov\footnote{e-mail: andrei@theory.caltech.edu}}
\\
\bigskip\bigskip
{\it California Institute of Technology 452-48,
Pasadena CA 91125 \\
\bigskip
and\\
\bigskip
Institute for Theoretical and 
Experimental Physics, \\
117259, Bol. Cheremushkinskaya, 25, 
Moscow, Russia}\\

\vskip 1cm
\end{center}

\begin{abstract}
It was conjectured recently that the string worldsheet theory 
for the fast moving string in AdS times a sphere becomes effectively  first
order in the time derivative and describes the continuous
limit of an integrable spin chain. In this paper we will
try to make this statement more precise. We interpret the first order
theory as describing the long term evolution of the tensionless
string perturbed by a  small tension.  The long term evolution
is a Hamiltonian flow on the moduli space of periodic trajectories.
It should correspond to the renormgroup flow on the field
theory side.  
\end{abstract}

\end{titlepage}

\section{Introduction.}
The AdS/CFT correspondence relates weakly
coupling limit of the Type IIB string theory to the
strongly coupled limit of the ${\cal N}=4$
Yang-Mills theory. 
It is hard to imagine that this type of
a correspondence would allow for quantitative
checks besides the comparison of the quantities
protected by the supersymmetry.
But the recent research revealed
several examples where some nontrivial parts of the
Yang-Mills perturbation theory are reproduced
from the string theory computations. 
The first work in this direction was  the 
computation of the expectation value of the circular Wilson
loop \cite{ESZ,DG}. It was followed by the discovery of
the BMN limit \cite{BFFHP,Metsaev,BMN} and the
``spinning string'' solutions which we will discuss in this
paper. In these computations supersymmetry alone is not enough
to guarantee the agreement of the results of the
string theory and the field theory.  
It turns out that in some field theory computations
the perturbation series depend on the coupling constant
$\lambda$ only in  the combination  
$\lambda/J^2$ where $J$ is a large integer.
If $J^2>>\lambda$  the perturbative computations
can presumably be trusted even when $\lambda$
is large, and when $\lambda$ is large they can be
matched with the string theory computations.  
At the moment there is no solid explanation of
why it  works, and even whether this is true
to all orders of the Yang-Mills perturbation theory
(see \cite{SerbanStaudacher} for one of 
the most recent discussions.)
But there are  several computations with the
impressive agreement between the field theory and the
string theory.

The ``spinning strings'' solutions were first considered 
in the context of the AdS/CFT correspondence  
in \cite{FT02,Tseytlin,Russo}.  
Various computations in the classical dynamics
of these solutions lead to the series in 
the small parameter which on the field theory
side is identified with $\lambda/J^2$. It was conjectured
 that the Yang-Mills
perturbation theory in the   corresponding sector
is reproduced by the classical dynamics of the spinning
strings. The corresponding Yang-Mills operators are the
traces of the products of the large number of
the elementary fields of the Yang-Mills theory;
$J$ corresponds roughly speaking to the number
of the elementary fields under the trace. 
The one-loop anomalous dimension of such operators
was computed in \cite{MZ,BS} and the perfect 
agreement was found with the classical string computations;
see the recent review \cite{TseytlinReview} for the
details. It turns out that the single trace operators
in the ${\cal N}=4$ Yang-Mills theory can be thought of
as quantum states of the spin chain, and the one loop 
anomalous dimension corresponds to the integrable Hamiltonian. 

A direct correspondence between the quasiclassical states 
of the spin chain and the classical string
solutions was proposed recently in \cite{Kruczenski}.
It was suggested that in the high energy limit the 
string worldsheet theory becomes effectively first order in 
the time derivative and  agrees with the Hamiltonian
evolution in the spin chain. In our paper we will try to 
generalize this statement and make it more precise.  

The characteristic property of the spinning strings, which
was first clearly explained in \cite{MateosMateosTownsend},
is that their worldsheets are nearly-degenerate.
In all the known situations when there is an agreement with
the field theory perturbative computation, every point of the
string is moving very fast, approaching the speed of light.
Therefore ``spinning strings'' are actually fast 
moving strings\footnote{Fast moving strings were also
considered in this context in \cite{AleksandrovaBozhilov}.}.
This observation suggests that there is a correspondence
between a certain class of the Yang-Mills operators
and the parametrized null surfaces in $AdS_5\times S^5$
\cite{SpeedingStrings}. A null surface is a surface with degenerate
metric, ruled by the light rays. A parametrized null surface
is a null surface $\Sigma$ with a function 
$\sigma:\Sigma\to S^1$ which is constant on the light rays.
(On the field theory side $\sigma$ can be thought of as
parametrizing ``the position of the elementary field operator
inside the trace''.)
A parametrized null surface can be specified
by the embedding functions $x(\sigma,\tau)$
with values in $AdS_5\times S^5$ such that for a fixed
$\sigma=\sigma_0$ the functions $x(\sigma_0,\tau)$ describe a 
light ray with the affine parameter $\tau$, and
$(\partial_{\sigma}x,\partial_{\tau}x)=0$. The embedding
functions are defined modulo the ``gauge transformations''
with the infinitesimal form
$\delta x= \phi(\sigma) \partial_{\tau}x$ where
$\phi(\sigma)$  is an arbitrary periodic function
of $\sigma$.

There is an interesting special
case when the null surface is generated by the orbits
of the lightlike Killing vector field $V$ in 
$AdS_5\times S^5$.
The corresponding field theory operators are characterized
by a special property that their engineering dimension
is equal to a certain combination of conserved charges. 
In this special case
the one loop anomalous dimension 
should be equal on the string theory
side to the value of the conserved charge corresponding
to $V$. We have shown in \cite{SpeedingStrings} that 
this charge is proportional to the following 
``action functional'':
\begin{equation}\label{CosetAction}
S[x]=\int_{S^1} d\sigma \; 
(\partial_{\sigma}x(\sigma,\tau), \partial_{\sigma}x(\sigma,\tau))
\end{equation}
with the coefficient of the order $\lambda/J^2$.
(This formula requires a choice of the closed contour on the null
surface, but the result of the integration does not actually depend 
on this choice.)
The definition of the special class of operators for which
the engineering dimension equals a combination of charges
makes sense for finite $\lambda\over J^2$.
What is special about the extremal surfaces corresponding
to this particular class of operators for finite $\lambda\over J^2$?
We describe this class of extremal surfaces in Section 2
to the first order in $\lambda\over J^2$.

We will also generalize the expression (\ref{CosetAction}) 
for the anomalous dimension for the case when $\Sigma$ is 
a general null-surface, not necessarily ruled by the orbits
of the symmetry. (The solutions of \cite{EMZ} belong to this
more general class.)
Following the idea of \cite{Kruczenski}
we will study the long term evolution of the approximating
nearly-degenerate extremal 
surface $\Sigma(\epsilon)$, $\epsilon^2=\lambda/J^2$, 
$\Sigma(0)=\Sigma$. 
We show that this long-term evolution is a 
Hamiltonian flow on the moduli space of the null-surfaces.
The generating function corresponds to the anomalous
dimension of the corresponding Yang-Mills operator.
The result is a very natural generalization of 
(\ref{CosetAction}):
\begin{equation}\label{GeneralCosetAction}
S[x]=
\int_0^{2\pi}d\tau\int_0^{2\pi} d\sigma\;
(\partial_{\sigma}x(\sigma,\tau), 
\partial_{\sigma}x(\sigma,\tau))
\end{equation}
This is a functional on the space of null-surfaces.
For its definition it is essential that all the light-like geodesics
in $AdS_m\times S^n$ are periodic. Therefore the null-surfaces are
also periodic, just like solutions of the massless field equations.
The integration over $\tau$ corresponds to taking the average
over the period, see Section \ref{SectionGeneral} for details.
The value of $S$ on the contour should correspond on the field theory 
side to the anomalous dimension of the corresponding operator. 

The null-surface perturbation theory was studied in a 
closely related context in \cite{dVGN}.

\vspace{5pt}
\noindent
\underline{The structure of the paper.}
In Section 2 we will study the perturbations
of the null surfaces  corresponding to the special class 
of operators for which the engineering
dimension is equal to a certain combination of
the $R$-charges.  On the AdS side this is reflected in 
the null-surface $\Sigma(0)$ being invariant under
the symmetry generated by the null Killing vector $V$. 
There are restrictions on the nearly-degenerate
worldsheet $\Sigma(\epsilon)$ 
following from the fact that the
operators of this class mix only
 among themselves under the renormgroup.
We show that these restrictions can be satisfied.

In Section 3 we will study the perturbations of the
null-surfaces which are not generated by the orbits of
the light-like Killing vector. 
We will describe the ``long-term''
or ``secular'' behavior of $\Sigma(\epsilon)$. 
The moduli space of  parametrized 
null-surfaces is a symplectic manifold, and the long-term evolution is a 
Hamiltonian flow 
corresponding to the renormgroup flow on the field theory side.

\section{Perturbation of the degenerate
surfaces ruled by the orbits of the light-like
Killing vector.}
\subsection{Summary of this section.}
Let $V$ be a lightlike Killing vector field in
$AdS_5\times S^5$. Consider the null surfaces
which are ruled by the orbits of $V$. These
null surfaces correspond to the Yang-Mills operators
of the form  $\trace F(X,Y,Z)$  
where $F(X,Y,Z)$ is some  (unsymmetrized) product
  of $X,Y,Z$; 
$X=\Phi_1+i\Phi_2$,
$Y=\Phi_3+i\Phi_4$ and $Z=\Phi_5+i\Phi_6$ are
the complex combinations of the scalar fields.
For these operators the charge corresponding
to $V$ is zero in the free theory.
Let $\cal O$ be an operator of this type.

What can we say about the extremal surface $\Sigma_{\cal O}$ 
corresponding to such an operator ${\cal O}$? 
We will argue that to the
first order in $\epsilon^2={\lambda\over J^2}$ the class  of 
extremal surfaces corresponding to this special 
type of operators can be characterized
as follows. For each point $x\in \Sigma_{\cal O}$ there
is a null-surface $\Sigma(0)$  ruled by the orbits of $V$ 
and such that in the vicinity of
$x$ the deviation of $\Sigma_{\cal O}$ from $\Sigma(0)$
is of the form:
\begin{equation}\label{Approximation}
	x(\tau,\sigma)=x_0(\tau,\sigma)+
	\epsilon^2\eta_1(\tau,\sigma)+\ldots
\end{equation}
where $\eta_1$ has the property:
\begin{equation}\label{VVeta}
	[V,[V,\eta_1]]=0
\end{equation}
Here $[V,\eta_1]=\nabla_{V}\eta_1-\nabla_{\eta_1}V$ denotes 
the commutator
of two vector fields; one of these fields is defined only
on the surface $\Sigma(0)$, therefore the commutator
is also defined only on $\Sigma(0)$. 
The property $[V,[V,\eta_1]]=0$ is what characterizes this 
special class 
of string worldsheets to the first order in $\epsilon^2$.

Unlike the null-surface $\Sigma(0)$, the nearly-degenerate
surface $\Sigma(\epsilon)$ is not invariant under $V$.
But we can describe the variation of $\Sigma(\epsilon)$
under $V$ rather explicitly. Indeed, we can see from 
(\ref{Approximation}) that the translation of
$x$ by $V$ with the infinitesimal parameter $\mu$ is:
\begin{eqnarray}
e^{\mu V.} x(\tau,\sigma)=
x(\tau,\sigma)+\mu\epsilon^2 [V,\eta_1](\tau,\sigma)=\\[5pt]
=(x_0(\tau,\sigma)+\mu\epsilon^2 [V,\eta_1](\tau,\sigma))
+\epsilon^2\eta_1(\tau,\sigma)
\end{eqnarray}
One can see that when the condition (\ref{VVeta}) is satisfied,
$x_0+\mu\epsilon^2 [V,\eta_1]$ determines the
infinitesimally deformed null-surface. Therefore the translation
of the nearly-degenerate extremal surface 
$\Sigma(\epsilon)$ by $V$ corresponds to the
deformation of the approximating null-surface $\Sigma(0)$. 

Let us formulate it more precisely.
Notice that a light-like Killing vector field $V$ in $AdS_5\times S^5$
	can always be represented as $V=U_A+U_S$ where
	$U_A$ and $U_S$ are Killing vector fields on $AdS_5$
	and $S^5$ respectively; $(U_A,U_A)=1$ and $(U_S,U_S)=-1$.
	Let $Q_S$ denote the 
	conserved charge corresponding to $U_S$.
Let ${\cal N}$
be the moduli space of parametrized null-surfaces ruled by the
orbits of $V$ and ${\cal M}_{J}$ be the moduli
space of extremal surfaces of the special type
characterized by Eq. (\ref{VVeta}) and such that $Q_S=J$.

Let us
choose some map $\Lambda: {\cal N}\to {\cal M}_{J}$,
such that:
\begin{enumerate}
	\item
		For any parametrized null-surface $\Sigma(0)$
		the image 
$\Lambda(\Sigma(0))$ is an extremal surface
deviating from $\Sigma(0)$ by the terms of the order 
$\epsilon^2={\lambda\over J^2}$
\item
	The density of $Q_S$
	on $\Sigma(\epsilon)=\Lambda(\Sigma(0))$ in the limit 
	$\epsilon\to 0$ is proportional to ${1\over\epsilon}d\sigma$
	where $\sigma$ is the parametrization 
	of the null-surface
	$\Sigma(0)$: 
	\begin{equation}\label{ChargeParametrization}
		\mbox{density of}\; Q_{S} \; ={\sqrt{\lambda}\over 4\pi}
		{1\over\epsilon}d\sigma +O(1)\;\;\;\;
		\mbox{when}\;\;	\epsilon\to 0
	\end{equation}
\end{enumerate}
	The action of $V$ on ${\cal M}_J$
by translations 
is conjugate by $\Lambda$ to some one-parameter
group of transformations of ${\cal N}$.
It turns out that this one-parameter group of transformations
to the first order in $\epsilon^2$ does not depend on the
choice of $\Lambda$. It has the following meaning
in the dual field theory.
We can identify ${\cal N}$ with the space of continuous
operators in the free field theory. Then the one-parameter
group of transformations which we described corresponds
to the renormgroup transformations of the continuous
operators when we turn on the interaction $\lambda/J^2$. 
This can be summarized in the commutative
diagramm:
\begin{equation}\label{CommutativeDiagramm}
\begin{array}{ccc}
\;\;\;\;\;\;\Sigma(0) & \stackrel{
\mbox{\small RG acting on null-surfaces}}
{\longminus\!\!\!\longminus\!\!\!\longminus\!\!\!\longminus
\!\!\!\longrightarrow} &
\widetilde{\Sigma}(0)\\[5pt]
\;\;\;\;\;\; \Lambda
\begin{array}{c}|\\[-5pt]|\\[-5pt] \downarrow\end{array}
\;\;
&&
 \!\!\!\!\Lambda
\begin{array}{c}|\\[-5pt]|\\[-5pt] \downarrow\end{array}
\\[5pt]
\;\;\;\;\;\;\Sigma(\epsilon) & \stackrel{
\mbox{\small shift by  Killing vector field}}
{\longminus\!\!\!\longminus\!\!\!\longminus\!\!\!\longminus
\!\!\!\longrightarrow} &
\widetilde{\Sigma}(\epsilon)
\end{array}
\end{equation}
In the rest of this section we will explain how to construct
the extremal surfaces satisfying the conditions (\ref{Approximation}),
(\ref{VVeta}).

\subsection{General facts about the nearly-degenerate
surfaces.}\label{GeneralFacts}
Consider the extremal surface in $AdS_5\times S^5$ which
is nearly-degenerate (close to being null).
Calculations are simplified with a special choice of 
the worldsheet coordinates:
\begin{eqnarray}\label{Constraints}
&&\left({\partial x\over \partial \tau},
{\partial x\over \partial\tau}\right)+
\epsilon^2 \left({\partial x\over\partial\sigma},
{\partial x\over\partial\sigma}\right)=0\\[5pt]
&&\left(
{\partial x\over \partial \tau},
{\partial x\over \partial\sigma}\right)=0 
\label{SigmaOrthogonalTau}
\end{eqnarray}
where $\epsilon$ is a small parameter measuring the
deviation of the worldsheet from a null-surface.
We assume that $\sigma$ is periodic with the period $2\pi$.
We choose the small parameter $\epsilon$ so that the 
embedding function $x(\tau,\sigma)$ has a finite limit when 
$\epsilon\to 0$. In this limit $x(\tau,\sigma)$ describes an 
embedding of the null-surface $x_0(\tau,\sigma)$. 
If we choose $\sigma$ as the parametrization of this limiting
null-surface then the density of $Q_S$ will agree with this
parametrization in the sense of Eq. (\ref{ChargeParametrization}).
The string worldsheet action is:
\begin{equation}\label{WorldsheetAction}
S={\sqrt{\lambda}\over 4\pi}\int d\sigma d\tau
\left[ {1\over\epsilon}(\partial_{\tau}x,\partial_{\tau}x)
-\epsilon (\partial_{\sigma}x,\partial_{\sigma}x)\right]
\end{equation}
The string equation of motion is:
\begin{equation}\label{SEM}
{1\over\epsilon}D_{\tau}\partial_{\tau}x-
\epsilon D_{\sigma}\partial_{\sigma}x=0
\end{equation}
We denote $D_{\tau}$ and $D_{\sigma}$ the worldsheet
covariant derivatives. They act on the vector-functions
on the worldsheet with values in the tangent space to
$AdS_5\times S^5$. The general definition is 
$$D_{\tau}\xi^{\mu}=
\partial_{\tau}\xi^{\mu}+
\Gamma^{\mu}_{\nu\rho}\partial_{\tau}x^{\nu}\xi^{\rho}$$
where $x^{\mu}=x^{\mu}(\tau,\sigma)$ are
the coordinate functions specifying the embedding of the
string worldsheet into the target space and
$\xi^{\mu}=\xi^{\mu}(\tau,\sigma)$ is a vector-function on
the worldsheet with values in the tangent space
$T(AdS_5\times S^5)$. Somewhat schematically, one can write
$D_{\tau}\xi^{\mu}=
\partial_{\tau}x^{\nu}\nabla_{\nu}\xi^{\mu}$ where
$\nabla_{\nu}$ is the covariant derivative in the tangent
bundle to the target space. More precisely, 
$D_{\sigma}$ is the natural connection in the ten-dimensional
vector bundle over the worldsheet which is the restriction
 to the worldsheet of the tangent bundle of $AdS_5\times S^5$. 
 This natural connection
 is induced from the Levi-Civita
 connection on $T(AdS_5\times S^5)$.

One can look for a solution  to (\ref{SEM}) 
as a power series
in $\epsilon^2$:
\begin{equation}
x(\sigma,\tau)=x_0(\sigma,\tau)+
\epsilon^2\eta_1(\sigma,\tau)+
\epsilon^4\eta_2(\sigma,\tau)+\ldots
\end{equation}
where $x_0(\sigma,\tau)$ is a null-surface.
The first deviation $\eta_1$ satisfies the inhomogeneous
Jacobi equation:
\begin{equation}\label{InhomogeneousJacobi}
{D^2\eta_1 \over \partial\tau^2}
+R\left({\partial x_0\over\partial\tau},\eta_1\right)
{\partial x_0\over\partial\tau} =
{D\over\partial\sigma}{\partial x_0\over\partial \sigma}
\end{equation}
and the constraints:
\begin{eqnarray}\label{ConstraintsOnEta}
&&(D_{\sigma}\eta_1 , \partial_{\tau}x_0)+
(D_{\tau}\eta_1 ,\partial_{\sigma}x_0)=0\label{FirstEtaConstraint}
\\[5pt]
&&(D_{\tau}\eta_1 ,\partial_{\tau}x_0)=
-{1\over 2}(\partial_{\sigma}x_0,
\partial_{\sigma}x_0)\label{SecondEtaConstraint}
\end{eqnarray}
where $R$ is the curvature tensor of the target space
(we will remind its definition in a moment).
The constraints (\ref{ConstraintsOnEta})
on $\eta$ follow from the constraints 
(\ref{Constraints}) on $x$. The inhomogeneous Jacobi 
equation\footnote{The Jacobi equation describes
the infinitesimal variation of a geodesic, see for 
example Appendix 1 of \cite{Arnold}. We decided
to keep this name for the equation (\ref{InhomogeneousJacobi})
which describes
the infinitesimal resolution of the null-surface becoming
an extremal surface. Indeed, 
the null-surface is composed of the null-geodesics.
After the resolution, these null-geodesics become
time-like curves. It is not true that these time-like curves
are geodesics, because there is a "driving force" 
$D_{\sigma}\partial_{\sigma}x$ on the right hand side of 
(\ref{InhomogeneousJacobi}). This driving force,
resulting from the tension of the string, makes the equation
inhomogeneous.} 
(\ref{InhomogeneousJacobi}) can be derived
from the equations of motion (\ref{SEM}) in the
following way. Consider the family of worldsheets
$\Sigma(\epsilon)$ parametrized by $\rho=\epsilon^2$.
This family of two-dimensional manifolds "sweeps"
some three-dimensional manifold (one boundary of this
three-dimensional manifold is the null-surface 
$\Sigma(0)$). 
Let us think of $\rho,\sigma,\tau$ as coordinates
on this three-dimensional manifold. 
Consider the equation (\ref{SEM}):
$D_{\tau}\partial_{\tau}x(\rho,\sigma,\tau)-
 \rho D_{\sigma}\partial_{\sigma}x(\rho,\sigma,\tau)=0$.
Differentiate it with respect to $\rho$:
\begin{equation}\label{DifferentiateSEM}
D_{\rho}D_{\tau}\partial_{\tau}x-
D_{\sigma}\partial_{\sigma}x-
\rho D_{\rho}D_{\sigma}\partial_{\sigma}x=0
\end{equation}
Now we have to take into account 
that the covariant derivatives do not commute.
They do not commute because the target space has a
non-zero Riemann tensor. 
 To define the Riemann tensor, one
takes two vector fields $\xi,\eta$ and computes  the
commutator of the covariant derivatives along these
two vector fields. The result is a section of
$\mbox{End}(T)(AdS_5\times S^5)$ --- the bundle
of linear maps from the tangent space to itself.
This section is a bilinear function of $\xi,\eta$ called
$R(\xi,\eta)$:
\begin{equation}\label{DefinitionOfR}
R(\xi,\eta)=-\nabla_{\xi}\nabla_{\eta}+
\nabla_{\eta}\nabla_{\xi}
+\nabla_{[\xi,\eta]}
\end{equation}
For given $\xi$ and $\eta$, $R(\xi,\eta)$ is a 
{\em matrix} acting in the tangent space 
to $AdS_5\times S^5$.  
The vector fields $\partial_{\rho}$, $\partial_{\sigma}$
and $\partial_{\tau}$ are defined only on the 
three-dimensional submanifold. But still, we can compute
their commutators and the commutators of the corresponding
covariant derivatives. We get, in particular, 
$[\partial_{\rho},\partial_{\tau}]=0$ and therefore
$$
[D_{\rho},D_{\tau}]=-R(\partial_\rho,\partial_\tau)
$$
Let us use this formula in (\ref{DifferentiateSEM}).
Taking into account also that $D_{\rho}\partial_{\tau}x=
D_{\tau}\partial_{\rho}x$ we get:
\begin{equation}
D_{\tau}D_{\tau}\partial_{\rho}x+
R(\partial_{\tau}x,\partial_{\rho}x)\partial_{\tau}x-
D_{\sigma}\partial_{\sigma}x-
\rho D_{\rho}D_{\sigma}\partial_{\sigma}x=0
\end{equation}
In this equation, let us put $\rho=0$. 
Since $\partial_{\rho}x|_{\rho=0}=\eta_1$
we get (\ref{InhomogeneousJacobi}).

Now we will consider the inhomogeneous
Jacobi equation in the special case
when $\Sigma(0)$ is ruled by the orbits of the light-like Killing
vector field. Our aim is to show that in this special case 
there are solutions satisfying (\ref{VVeta}).

\subsection{A special case of 
the inhomogeneous Jacobi equation.}
We will start by rewriting (\ref{InhomogeneousJacobi})
in the special case when $\Sigma(0)$ is ruled by
the orbits of $V$, that is $\partial_{\tau}x_0=V(x_0)$:
\begin{equation}\label{Jacobi}
{D^2\eta\over \partial \tau^2}+R(V,\eta)V =
{D\over\partial\sigma}
{\partial x_0\over\partial \sigma}
\end{equation}
Let us introduce an abbreviation for the covariant derivative;
for two vector fields $\alpha$ and $\beta$ we will denote
$\alpha.\beta^{\mu}=\alpha^{\nu}\nabla_{\nu}\beta^{\mu}$.
Taking into account (\ref{DefinitionOfR}) we have:
\begin{eqnarray}
&&D_{\tau}\eta=V.\eta=[V,\eta]+\eta.V\\[5pt]
&&D_{\tau}^2\eta=V.(V.\eta)=V.[V,\eta]+[V,\eta].V
-R(V,\eta)V=\\[5pt]&&=[V,[V,\eta]]+2[V,\eta].V-R(V,\eta)V
\end{eqnarray}
This  allows us to rewrite (\ref{Jacobi}) as:
\begin{equation}\label{CommutatorFormA}
[V,[V,\eta]]+2[V,\eta].V=
{D\over\partial\sigma}{\partial x_0\over\partial \sigma}
\end{equation}
Since $V$ is a Killing field, its covariant derivative
is antisymmetric: $\nabla_{\mu}V_{\nu}=-\nabla_{\nu}V_{\mu}$.
Therefore for any vector field $\alpha$
we can write $\alpha.V=\iota_{\alpha}\omega$
where $\omega_{\mu\nu}=\nabla_{\mu}V_{\nu}$
is a closed two-form. With this notation 
Eq. (\ref{CommutatorFormA}) becomes:
\begin{equation}\label{CommutatorForm}
[V,[V,\eta]]+2\iota_{[V,\eta]}\omega=
{D\over\partial\sigma}{\partial x_0\over\partial \sigma}
\end{equation}
The null-surfaces ruled by the orbits of the null Killing
correspond to  operators of the form
$\trace F(X,Y,Z)$.  
Consider a degenerate surface $\Sigma(0)$ generated by
the orbits of $V$ and its deformation $\Sigma(\epsilon)$
corresponding to turning on the coupling constant.
 Although $\Sigma(0)$ is invariant under
$V$, its deformation $\Sigma(\epsilon)$ is not invariant. Let us
consider the translation of $\Sigma(\epsilon)$ by the vector
field $V$ with the parameter $\mu$, schematically 
$e^{\mu V}.\Sigma(\epsilon)$. This corresponds to the action of the
renormgroup on the operator in the theory with a finite coupling
constant. 
The operators of the type $\mbox{tr}\;F(X,Y,Z)$ are only 
mixing among themselves under the renormgroup at the level
of one loop. This implies that
the translation along $V$ of the deformation of the 
null-surface ruled by the orbits of $V$ 
should be the deformation
of some other null-surface which is also ruled by the
orbits of $V$. 
For the infinitesimal deformation this means that
\begin{equation}\label{ZeroDoubleCommutator}
[V,[V,\eta_1]]=0
\end{equation}
Indeed $\mu \epsilon^2 [V,\eta_1]$ is 
the variation of the deformed worldsheet under the shift
by $e^{\mu V.}$. Then the condition $[V,[V,\eta_1]]=0$ 
implies that: 
\begin{enumerate}
\item{$[V,\eta_1]$ is  a solution of the
{\em homogeneous} Jacobi equation and therefore
$x_0+\mu \epsilon^2 [V,\eta_1]$ can be considered
as defining the deformed null-surface\footnote{That 
this deformed surface is degenerate
 follows from the constraints 
(\ref{FirstEtaConstraint}),(\ref{SecondEtaConstraint}) 
and from $V$ being a Killing vector.}}
\item{This deformed null surface is again ruled by the 
orbits of $V$.}
\end{enumerate}
Therefore under the condition (\ref{ZeroDoubleCommutator})
the shift of $\Sigma(\epsilon)$ by $V$ can be ``compensated''
by the deformation of $\Sigma(0)$, and the deformed
$\Sigma(0)$ is again ruled by the orbits of $V$.
This is precisely the statement that the
diagramm (\ref{CommutativeDiagramm}) is commutative, to the
first order in $\epsilon^2$.

Can we find $\eta_1$ satisfying (\ref{CommutatorForm}) 
and (\ref{ZeroDoubleCommutator})?
It turns out that we can. Indeed, with the condition
(\ref{ZeroDoubleCommutator}) Eq. (\ref{CommutatorForm}) 
becomes:
\begin{equation}\label{HamiltonianForm}
2\iota_{\zeta}\omega=
{D\over\partial\sigma}{\partial x_0\over\partial \sigma}
\end{equation}
where we denoted
$$
\zeta=[V,\eta_1]
$$ 
We want to study the space
of solutions of the equation (\ref{HamiltonianForm}).
The 2-form $\omega$ is degenerate, therefore we have
to make sure that the right hand side
of (\ref{HamiltonianForm}) belongs to the image of
$\omega$. 
To describe the kernel of $\omega$
we decompose $V=V_{AdS_5}+V_{S^5}$.
Here $V_{AdS_5}$ is the component of $V$ in the tangent
space to $AdS_5$ and $V_{S^5}$ is the component in the
tangent space to $S^5$. 
The kernel of $\omega$ is generated by $V$
and $\tilde{V}=V_{AdS_5}-V_{S^5}$.
Notice that 
$D_{\sigma}\partial_{\sigma}x_0$ is orthogonal to 
$V$ (the proof of this fact uses that $V$ is a Killing
vector and $V$ is orthogonal to $\partial_{\sigma}x_0$).
Therefore it is orthogonal to one of the vectors in
the kernel of $V$.  It does not follow that
$D_{\sigma}\partial_{\sigma}x_0$ is orthogonal
to $\tilde{V}$. But remember that $\partial_{\sigma}x_0$
is defined modulo $V$. Adding to $\partial_{\sigma}x_0$
something proportional to $V$ we can make it orthogonal
to $\tilde{V}$. Indeed, we have 
\begin{equation}
(\tilde{V},D_{\sigma}\partial_{\sigma}x_0)
= \partial_{\sigma}
(\tilde{V},\partial_{\sigma}x_0) 
\end{equation}
and one can change $x_0$ to $\tilde{x}_0$ where
\begin{equation}
\partial_{\sigma}\tilde{x}_0=\partial_{\sigma} x_0
-\left({(\tilde{V},\partial_{\sigma}x_0)-C\over 
(\tilde{V},V)}\right)V
\end{equation}
where $C$ is a constant. We adjust $C$ so that 
$\tilde{x}_0$ is periodic. We have 
$(\tilde{V},\partial_{\sigma}\tilde{x}_0)=C$. Now
$(\tilde{V},D_{\sigma}\partial_{\sigma}\tilde{x}_0)=0$
and therefore 
$D_{\sigma}\partial_{\sigma}\tilde{x}_0$
is orthogonal to the kernel of $\omega$
and therefore $\omega$ is invertible on it.

We have to also take care of the constraints 
(\ref{FirstEtaConstraint}), (\ref{SecondEtaConstraint}).
Notice that $\zeta=[V,\eta_1]$ is determined
from (\ref{HamiltonianForm}) only up to 
a linear combination of $V$ and $\tilde{V}$.
The coefficient of $V$ is undetermined and corresponds
to the $\sigma$-dependent rescaling of the affine
parameter on the light ray.
The coefficient of  $\tilde{V}$ is fixed 
to satisfy (\ref{SecondEtaConstraint}). After that
$[V,\eta_1]$ is completely fixed modulo $V$.
It remains to satisfy (\ref{FirstEtaConstraint}).
Let us rewrite (\ref{FirstEtaConstraint}) in the following
form:
\begin{equation}
(V.\eta_1,\partial_{\sigma}x_0)+(D_{\sigma}\eta_1,V)=
([V,\eta_1],\partial_{\sigma}x_0)+
(D_{\sigma}\eta_1,V)-\omega(\partial_{\sigma}x_0,\eta_1)=0
\end{equation}
We can look for $\eta_1(\tau=0,\sigma)$ in the form
$\eta_1|_{\tau=0}=\alpha(\sigma) \tilde{V}+\beta(\sigma)$ 
where  $\beta$ is a vector orthogonal to both $V$ and
$\tilde{V}$ and $\alpha$ is a function of $\sigma$ such that:
$$
2 \partial_{\sigma}\alpha=-([V,\eta_1],\partial_{\sigma}x_0)
+\omega(\partial_{\sigma}x_0,\beta)
$$
There is a freedom in the choice of $\beta$,
the only constraint is that $\alpha$ determined
from this equation should be a periodic function
of $\sigma$. This is the freedom
to add to $\eta_1$ a constant vector $\Delta\eta_1$ 
(constant means $[V,\Delta\eta_1]=0$)  satisfying
$(D_{\sigma}\Delta\eta_1,V)+
(V.\Delta\eta_1,\partial_{\sigma}x_0)=0$. 
This corresponds to the $\epsilon^2$-deformation
of the null-surface remaining the null-surface. 

The solutions of 
(\ref{CommutatorForm}) which
have $[V,[V,\eta_1]]\neq 0$ 
correspond to operators of the
form ${\cal O}+{\lambda\over J^2}\widetilde{\cal O}$
where $\widetilde{\cal O}$ is not annihilated
by the symmetry corresponding to $V$.

\subsection{Example: the two-spin solution.}
Here we will consider as an example 
the two-spin solution of \cite{AFRT}. This solution is of the
type considered in this section, the corresponding null-surface
is $V$-invariant. We will reproduce the terms of the order
$\epsilon^2$ in the expansion of the worldsheet near the 
null-surface.

Let us parametrize the sphere $S^5$ by the three complex coordinates
$Y_I=x_I e^{i\phi_I}$ with $\sum_{I=1}^3 x_I^2=1$. Of the AdS space
we will need only a timelike geodesic, which we parametrize 
by $t$. The metric is $-dt^2+\sum |dY_I|^2$. 
The lightlike Killing vector is 
$$V={\partial\over\partial t}+
\sum_{I=1}^3 {\partial\over\partial\phi_I}$$
Consider the following null-surface $x_0^{\mu}(\sigma,t)$:
\begin{equation}\label{null}
x_I=x_I(\sigma),\;\;\; \phi_I(t)=t
\end{equation}
The one-form $g_{\mu\nu}V^{\nu}$ is\footnote{we denote the one-form
corresponding to the vector $V$ by the same letter; this should not 
lead to a confusion} $V=-dt+\sum x_I^2 d\phi_I$, therefore
$\omega=\sum dx_I^2\wedge d\phi_I$. 
 For any  vector $\xi$ we have 
$
\iota_{\xi}\omega={1\over 2}\sum \left[(\xi. x_I^2)d\phi_I -
(\xi.\phi_I) d x_I^2\right]
$. The one-form on the right hand side of 
(\ref{HamiltonianForm}) is:
\begin{equation}
D_{\sigma}\partial_{\sigma}x=\sum_I
(D_{\sigma}\partial_{\sigma} x_I) dx_I
\end{equation}
Eq. (\ref{HamiltonianForm}), together with the constraint
$(V,[V,\eta])=-{1\over 2}(\partial_{\sigma}x)^2$ 
can be solved as follows:
\begin{equation}
[V,\eta]={1\over 2}(\partial_{\sigma}x)^2
{\partial\over\partial t}-
{1\over 2} \sum_I x_I^{-1} D_{\sigma}\partial_{\sigma}x_I
{\partial\over\partial \phi_I}
\end{equation}
This means, that on the initial surface 
(\ref{null}) $\eta$ is a linear function of $t$:
\begin{equation}
\eta=t\left[
{1\over 2}(\partial_{\sigma}x)^2
{\partial\over\partial t}-
{1\over 2} \sum_I x_I^{-1} D_{\sigma}\partial_{\sigma}x_I
{\partial\over\partial \phi_I}
\right]
\end{equation}
Let us compare this  to the solution of \cite{AFRT}.
The solutions of \cite{AFRT} correspond to a special 
finite-dimensional subspace
in the space of null-surfaces, such that
the contours $x(\sigma,\tau)|_{\tau=\tau_0}$ are the periodic
trajectories of the C.~Neumann integrable system:
\begin{equation}
D_{\sigma}\partial_{\sigma} x_I=-w_I^2 x_I +x_I \sum w_J^2 x_J^2
\end{equation}
On such contours, 
\begin{equation}
\eta=t\left[
{1\over 2}((\partial_{\sigma}x)^2+\sum w_J^2 x_J^2)
{\partial\over\partial t}+
{1\over 2} \sum_I w_I^2
{\partial\over\partial \phi_I}
\right]\;\;\;\mbox{mod}\;V
\end{equation}
The expression $\kappa^2=\sum
(\partial_{\sigma}x_I)^2+\sum w_J^2 x_J^2$ is twice the energy of
the Neumann system. 
One can see that  $x_0+\epsilon^2\eta$ gives the zeroth and the 
first  terms in the expansion of the solution 
of Section 2.1 of \cite{AFRT} around 
the null-surface\footnote{
There is a difference in notations:
$
\kappa^2_{[AFRT]}=1+\epsilon^2\kappa^2,
\;\;
w^2_{I[AFRT]}=1+\epsilon^2 w^2
$}.

\section{The general case: $V$ is not a Killing vector field.}
\label{SectionGeneral}
In $AdS_5\times S^5$ the null-geodesics are all periodic with
the same period, in a sense that all the light rays emitted
from the given point in the future direction will refocus
in the future at some other point. 
This implies that  the null-surfaces in $AdS_5\times S^5$
are all periodic with the same integer period. 
The null-surfaces should correspond to the large charge
operators at zero coupling; the periodicity
of the null-surface corresponds to the fact that the
operators in the free theory  have zero anomalous dimension.  

Turning on a small coupling constant corresponds to considering
the extremal surfaces which are very close to being null.
Such surfaces are the worldsheets of the ``ultra-relativistic''
strings. Naively one could think that the
extremal surfaces which are close to the null-surfaces
are periodic modulo small corrections. 
But this is not true \cite{Kruczenski}. 
It turns out that the worldsheet of the
ultrarelativistic string is close to the degenerate
surface only locally, in the following sense.
For each point on the worldsheet 
 there is a neighborhood  with
the coordinate size of the  order the AdS radius where
the surface is indeed close to some null-surface.
But as we follow the time evolution 
the deviation of the extremal surface
from the null-surface accumulates in time, 
and eventually becomes of the order of the radius of the AdS
space. This is a manifestation of the
general phenomenon which is known in classical mechanics 
as the ``secular evolution'' or the ``long-term evolution''
of the perturbed integrable systems \cite{Arnold}.
If the string worldsheet was originally
close to a null-surface $\Sigma(0)$ then after evolving
for a period of time $\Delta T\sim \epsilon^{-2}$ it will
be close to some other null-surface 
$\Sigma(0)^{(\epsilon^2\Delta T)}$
which is different from $\Sigma(0)$. Therefore we
get a one-parameter family  of transformations on the
moduli space of the null-surfaces with the parameter 
$\Delta T$, or rather $\epsilon^2\Delta T$.
We call these transformations the ``long term evolution''
of the null-surfaces.
In fact the fast moving string determines a null-surface
and its parametrization, therefore we have a family
of transformations on the moduli space of parametrized 
null-surfaces.

Before we proceed with the analysis of the string, we outline
a general situation when this slow evolution is usually
found. 
Suppose that we have an integrable system on the phase
space $M$ with the
Hamiltonian $H_0$, and $H_0+\epsilon^2\Delta H$ is
a perturbed  Hamiltonian. We are interested in the
special case when the phase space $M$ has
a submanifold $M_T\subset M$ closed under the flow
of $H_0$, such that $H_0|_{M_T}$ is constant and 
all the trajectories of $H_0$
on $M_T$ are periodic  with the same
period $T$. Also, 
we require that the perturbation is such that 
the trajectories of $H_0+\epsilon^2\Delta H$
 which started near $M_T$ will  stay near $M_T$ at least 
on the time intervals $\Delta t\sim \epsilon^{-2}$.
In other words, the trajectory of the perturbed
Hamiltonian which started on $M_T$ should
be always close to some ``approximating'' 
periodic trajectory of  the unperturbed system. 
(This does not follow from anywhere; it is an additional
assumption which has to be verified.) 
The ``approximating'' periodic trajectory will
slowly drift. Let us calculate the velocity of the
drift. Suppose that we started at the point $x_0\in M_T$
on  the periodic
trajectory of $H_0$ with the period $T$. 
Let us denote $x_0(\tau)$ the periodic 
trajectory of $H_0$ starting at $x_0$.
The perturbation
drives us away from this periodic trajectory. Take $n$ an integer,
$n<<\epsilon^{-2}$.
After the time interval $nT$ we are close to the original
point $x_0$. The deviation from $x_0$ is:
\begin{equation}
\delta x=\epsilon^2
\int_0^{nT} d\tau\; \left(e^{(nT-\tau)H_0}\right)_*\omega^{-1}
d(\Delta H)(x_0(\tau))+o(\epsilon^2)
\end{equation}
Here $\left(e^{(nT-\tau)H_0}\right)_*$ denotes the
 translation of the vector in the tangent space
to $M$ at the point $x_0(\tau)$ forward 
to the point $x_0(nT)=x_0$ by the flow of $H_0$.
Let us compute $\iota_{\delta x}\omega$:
\begin{equation}\label{OmegaOnDeltaX}
\iota_{\delta x}\omega=
\epsilon^2 \left[\int_0^{nT} d\tau \; 
\left(e^{-(nT-\tau)H_0}\right)^* 
d\; \Delta H(x_0(\tau))\right]
+o(\epsilon^2)
\end{equation}
Because of our assumption the component of $\delta x$ which
is transverse to $M_T$ does not accumulate in time.
This means that for sufficiently large $n$ we have
${1\over n}\delta x$ approximately tangent
to $T_{x_0}M_T$ (the component transverse to $T_{x_0}M_T$ is
of the order ${\epsilon^2\over n}$.)
The one-form on the right
hand side of  (\ref{OmegaOnDeltaX}) simplifies if we
restrict it 
to the tangent space to $M_T$. If we take $\xi\in T_{x_0}M_T$
and compute $\omega(\delta x,\xi)$, we will get the difference
of $\epsilon^2\int_0^{nT}\Delta H= 
n\epsilon^2 \overline{\bf \Delta H}$ on the periodic
trajectory going through $x_0+\xi$ and the periodic
trajectory going through $x_0$. In this sense,
\begin{equation}
\iota_{\delta x}\omega|_{T_{x_0}M_T}=
n\epsilon^2 \; d\;
\overline{\bf \Delta H}
\end{equation}
We have the following picture.
 Consider the restriction of $\omega$ on $M_T$.
Because $H_0|_{M_T}=\mbox{const}$ the tangent vector
to the periodic trajectory is in the kernel
of $\omega|_{M_T}$. This means that $\omega|_{M_T}$
defines a closed two-form on the space of periodic
trajectories with the period $T$, which we will denote
${\bf \Omega}$. The ``averaged'' Hamiltonian
$\overline{\bf \Delta H}$ is a function on this space
of periodic trajectories. The secular evolution
is the vector field $\boldxi$ on the space of
periodic trajectories which satisfies
\begin{equation}
\iota_{\boldxi}{\bf \Omega}=d\;\overline{\bf \Delta H}
\end{equation}
In the rest of this section we will apply this general 
scheme to the ultrarelativistic string in
$AdS_5\times S^5$.

\subsection{Hamiltonian approach to the fast moving strings.}
Consider the fast  moving string in $AdS_5\times S^5$.
As explained in Section 2.2 of \cite{SpeedingStrings}
we can parametrize the worldsheet by the coordinates
$\sigma$ and $\tau$ such that the embedding functions
satisfy the constraints:
\begin{eqnarray}\label{ConstraintsA}
&&(\partial_{\tau}x,\partial_{\tau}x)+
\epsilon^2(\partial_{\sigma}x,\partial_{\sigma}x)=0\\[5pt]
&&(\partial_{\tau}x,\partial_{\sigma}x)=0
\label{ConstraintsB}
\end{eqnarray}
These conditions do not completely fix
$\sigma$ and $\tau$. They are preserved by the
infinitesimal reparametrizations of the following form:
\begin{equation}\label{ResidualCoordinateTransformations}
\delta_{(f_L,f_R)}x=
[f_L(\sigma+\epsilon\tau)+f_R(\sigma-\epsilon\tau)]
{\partial x\over\partial\tau}+
\epsilon [f_L(\sigma+\epsilon\tau)-f_R(\sigma-\epsilon\tau)]
{\partial x\over\partial\sigma}
\end{equation}
We will assume that $x$ is a series in even powers
of $\epsilon$: $x=x_0+\epsilon^2\eta_1+\epsilon^4\eta_2+\ldots$;
this form of $x$ is preserved by the transformations 
(\ref{ResidualCoordinateTransformations}) with 
$$f_L=f_0+\epsilon f_1 +\epsilon^2 f_2+\ldots$$
$$f_R=f_0-\epsilon f_1 +\epsilon^2 f_2-\ldots$$
Using this residual freedom in the choice of the
coordinates we can impose the following condition
on the projection of the string worldsheet on $S^5$:
\begin{eqnarray}
&&(\partial_{\tau}x_{S^5},
 \partial_{\sigma}x_{S^5})=C+O(\epsilon^2) 
\label{SpecialChoiceOfTilt}
\\[5pt] 
&&(\partial_{\tau}x_{S^5},\partial_{\tau}x_{S^5})+
\epsilon^2(\partial_{\sigma}x_{S^5},\partial_{\sigma}x_{S^5})
=-1+\tilde{C}\epsilon^2+O(\epsilon^4)
\label{SpecialChoice}
\end{eqnarray}
where $C$ and $\tilde{C}$ are both constants 
(do not depend on $\sigma$.) Rescaling $\epsilon$ 
and $\tau$ by $\epsilon^2\to (1-\tilde{C}\epsilon^2)\epsilon^2$
and $\tau\to (1-\tilde{C}\epsilon^2)^{-1/2}\tau$ we can put
\begin{equation}
\tilde{C}=0
\end{equation}
The initial conditions (\ref{SpecialChoiceOfTilt}),
(\ref{SpecialChoice}) are preserved by the equation of motion
$D_{\tau}\partial_{\tau}x-
\epsilon^2 D_{\sigma}\partial_{\sigma}x=0$.
This particular
choice of the coordinates  simplifies  the calculations.

In the limit $\epsilon=0$ the worldsheet of the string becomes
a collection of non-interacting massless particles.
This limiting system can be described by the action
\begin{equation}\label{Unperturbed}
S_0={1\over 2}\int d\tau\int_0^{2\pi}
d\sigma \left({\partial x\over\partial\tau},
{\partial x\over \partial\tau}\right)
\end{equation}
which is the first term of (\ref{WorldsheetAction}).
(In this section we will omit the overall coefficient
${\sqrt{\lambda}\over 4\pi}{1\over\epsilon}$ in front 
of the action.)
Introduction of $\epsilon>0$ corresponds to the
perturbation of this system by the interaction
between particles, which is described by the second term
on the right hand side of (\ref{WorldsheetAction}).
 The interaction term is 
\begin{equation}\label{Perturbation}
\Delta S = {1\over 2}\;\epsilon^2 \int d\tau
\int_0^{2\pi} d\sigma \left({\partial x\over\partial \sigma},
{\partial x\over\partial \sigma}\right)
\end{equation}
Let us reformulate this problem in the Hamiltonian
approach. We will begin with the study of the unperturbed system
(\ref{Unperturbed}). Consider first the
$S^n$ part. The unperturbed system can be thought of 
as a continuous
family of free non-interacting particles moving on a sphere.
For every fixed $\sigma=\sigma_0$, $\;x(\tau,\sigma_0)$ describes the
motion of a free particle which is  independent
of particles corresponding to other $\sigma\neq\sigma_0$.
The momentum conjugate to $x\in S^n$
is $p={\partial x\over\partial\tau}$, and the Hamiltonian
is $H_0={1\over 2}(p,p)$. This system is integrable.
For every $\sigma$ the corresponding point
of the string moves on its own geodesic in $S^n$, different geodesics
for different values of $\sigma$, and the velocity  
generally speaking may also depend on $\sigma$.
The geodesics in $S^n$ are periodic. We can parametrize
every geodesic by an angle $\psi\in [0,2\pi]$. 
For each $\sigma$ the ``angle'' variable $\psi(\sigma)$ satisfies 
$\partial_{\tau}\psi(\sigma,\tau)=f(\sigma)$
where $f(\sigma)$ is the $\sigma$-dependent frequency.
We want to study the effect of the small perturbation
(\ref{Perturbation}). Let us first introduce some
useful notations.

\vspace{10pt}
\noindent
\underline{Particle on a sphere.}
We will consider two symplectic manifolds. The first
is the phase space of a free particle moving on a sphere
with the Lagrangian $(\dot{x},\dot{x})$; we will denote
it $M$. This is the cotangent bundle of the sphere 
$M=T^*S^n$.
The second symplectic manifold is the moduli
space of the geodesics in $S^n$; we will call it $G$.
The natural symplectic form on $G$ can be constructed
in the following way. Let us parametrize each geodesic
by an angle $\psi$; we have 
$(\partial_{\psi}x,\partial_{\psi}x)=1$. 
The tangent space to the moduli space
of geodesics at a given geodesic is given by the
Jacobi vector fields $\xi$ which satisfy the Jacobi equation
$D_{\psi}^2 \xi - R(\partial_{\psi}x,\xi)\partial_{\psi}x=0$.
Given two Jacobi vector fields $\xi_1$ and $\xi_2$ we
define the symplectic 
form:
\begin{equation}\label{SymplecticFormOnJacobi}
{\bf\Omega}(\xi_1,\xi_2)=-(\xi_1, D_{\psi}\xi_2)+
(D_{\psi}\xi_1,\xi_2)
\end{equation}
The right hand side is evaluated at a particular
point on the geodesic (at some particular $\psi$.)
But it does not depend on the choice
of this point (because of the Jacobi equation).  
 It is closed because it is
actually a differential of the one-form
$(\partial_{\psi}x,\xi)$; this one-form does depend on the
choice of a point on a geodesic, but its differential
does not. Also, a ``trivial'' Jacobi field
$\xi_2=\partial_{\psi}x$ corresponding to the shift along
the geodesic is in the kernel of $\bf\Omega$. Indeed, 
$$
{\bf\Omega}(\xi, \partial_{\psi}x)=
(D_{\psi}\xi,\partial_{\psi}x)=0 
$$
because $(\partial_{\psi}x,\partial_{\psi}x)=1$
for both the original geodesic and its infinitesimal
deformation by the Jacobi field $\xi$.
Therefore
$\bf\Omega$ is a well defined two-form on the moduli
space of geodesics.

Consider the subspace $M_{\times}\subset M$ of the phase
space  where the velocity of the particle is nonzero. 
It is a fiber bundle over the moduli space of
geodesics $G$. Indeed, the position and the
velocity of the particle uniquely determines the geodesic on which the
particle is moving. This defines a projection map: 
\begin{equation}
\boldpi: M_{\times}\to G 
\end{equation}
from the phase space of the
particle to the moduli space of geodesics. 
We will try to use boldface
letters to denote  objects on $G$ to distinguish
them from the functions and forms on $M$. 
We decided
to use a boldface to denote the projection map 
because it takes values in $G$, so ${\boldpi}(p,x)$ determines
a point in $G$. The fiber  of $\boldpi$ is
$S^1\times {\bf R}_{\times}$ where ${\bf R}_{\times}$ is a 
real line without zero.
 The $S^1$ parametrizes the position $\psi$ on the
geodesic and ${\bf R}_{\times}$ determines the velocity 
$f=\sqrt{E}$ where we denoted $E=(p,p)$.
Let us introduce the 1-form ${\cal D}\phi$ on $M_{\times}$:
\begin{equation}
{\cal D}\phi={(p,dx)\over (p,p)}
\end{equation}
It is characterized by the properties: 1) the restriction
of ${\cal D}\phi$ on the fiber $S^1\times {\bf R}_{\times}$
is $E^{-1/2}d\psi$ where $\psi$ is the angle on $S^1$ and 
2) it is zero on any vector in $TM_{\times}$ having a projection
on  $TS^n$ orthogonal to $p$. 
 For a vector ${\bf v}\in TG$
we will define a lift ${\boldpi}^{-1}{\bf v}$ as a vector in
$TM_{\times}$ with ${\boldpi}_*({\boldpi}^{-1}{\bf v})={\bf v}$ 
and $dE(\boldpi^{-1}{\bf v})=0$ and 
${\cal D}\phi(\boldpi^{-1}{\bf v})=0$. 
This determines the connection
on the fiber bundle $M_{\times}\to G$.

The symplectic form on
$M_{\times}$ can be written in terms of ${\cal D}\phi$ and
the pull-back of the symplectic form on $G$:
\begin{equation}\label{SymplecticForm}
\omega={1\over 2}dE\wedge {\cal D}\phi +\sqrt{E}\; \boldpi^*{\bf\Omega}
\end{equation}
\underline{Particle on $AdS_m\times S^n$.}
It is straightforward to write the analogue of (\ref{SymplecticForm})
for the particle moving on $AdS_m$ and on $AdS_m\times S^n$.
We consider $AdS_m\times S^n$ with the metric of the mostly negative
signature (that is, the metric on $S^n$ is considered negative 
definite.) For two vectors  $\xi,\eta$ in the tangent space
to $AdS_m\times S^n$ we denote $(\xi,\eta)_A$ the scalar product
of their $AdS_m$ components, and $(\xi,\eta)_S$ the scalar
product of their $S^n$ components. In general,  the index $A$
will denote objects on $AdS_m$ and the index $S$ objects on the
sphere. 
Let us introduce the notations:
\begin{eqnarray}
&&E_A=(p,p)_A, \;\;\;\;\; E_S=(p,p)_S,\\[5pt]
&&{\cal D}\phi_A=E_A^{-1}(p,dx)_A,\;\;\; 
{\cal D}\phi_S=E_S^{-1}(p,dx)_S
\end{eqnarray}
(Notice that  
$E_A$ is positive and $E_S$ is negative.) We have
\begin{eqnarray}
\boldpi^*{\bf\Omega}_A={(dp\wedge dx)_A\over E_A^{1/2}}-
{(p,dp)_A\wedge (p,dx)_A\over E_A^{3/2}}\nonumber\\[5pt]
\boldpi^*{\bf\Omega}_S={(dp\wedge dx)_S\over (-E_S)^{1/2}}+
{(p,dp)_S\wedge (p,dx)_S\over (-E_S)^{3/2}}\nonumber
\end{eqnarray}
Therefore
\begin{equation}
\omega={1\over 2}dE_A\wedge {\cal D}\phi_A+
{1\over 2}dE_S\wedge {\cal D}\phi_S +
\sqrt{E_A}\boldpi^*{\bf\Omega}^*_A+
\sqrt{-E_S}\boldpi^*{\bf\Omega}^*_S
\end{equation}
Here $\boldpi^*{\bf\Omega}^*_A$ and $\boldpi^*{\bf\Omega}^*_S$
are lifted from the moduli space of geodesics
on $AdS_m$ and $S^n$, respectively; \hspace{5pt}
${\cal D}\phi_S={(p,dx)_S\over (p,p)_S}$.

\vspace{10pt}
\noindent 
\underline{String on $AdS_m\times S^n$.}
 Let us proceed with our original system, which is
a continuous family of free particles. 
The phase space of the system is the ``loop space''
$LM$ which consists of the contours $(p(\sigma),x(\sigma))$
satisfying the constraints $(p,\partial_{\sigma}x)=0$
and 
$(p,p)+\epsilon^2(\partial_{\sigma}x,\partial_{\sigma}x)=0$. 
 The symplectic form is an integral over  $\sigma$:
\begin{eqnarray}\label{OmegaOnLM}
\omega=\int d\sigma\;\left[
{1\over 2}dE_A(\sigma)\wedge {\cal D}\phi_A(\sigma)+
{1\over 2}dE_S(\sigma)\wedge {\cal D}\phi_S(\sigma)+
\right.\nonumber\\[5pt]
\left.+\sqrt{E_A(\sigma)}\boldpi^*{\bf\Omega}^*_A(\sigma)+
\sqrt{-E_S(\sigma)}\boldpi^*{\bf\Omega}^*_S(\sigma)
\right]
\end{eqnarray}
We want to derive an evolution equation on $LG$.
We use the boldface  for the objects living on $G$ or $LG$, 
therefore our goal is to arrive at the equation where
all the letters are bold.
 The differential of the perturbation Hamiltonian is 
$$d\;\Delta H=\int d\sigma\; 
(\partial_{\sigma}x,D_{\sigma}dx)=
-\int d\sigma (D_{\sigma}\partial_{\sigma}x,dx)$$
Let us decompose $dx$ as the sum of the component parallel to 
$p=\partial_{\tau}x$
and the component orthogonal to $p$.
We get:
\begin{eqnarray}
&&d\;\Delta H=
\int d\sigma\left[
-(p(\sigma),D_{\sigma}\partial_{\sigma}x)_A
{\cal D}\phi_A(\sigma)-\right. \nonumber\\[5pt]
&&\left. 
-(p(\sigma),D_{\sigma}\partial_{\sigma}x)_S
{\cal D}\phi_S(\sigma) -\left(dx(\sigma),
(D_{\sigma}\partial_{\sigma}x)_{\perp}\right)
\right]
\end{eqnarray}
Here $(D_{\sigma}\partial_{\sigma}x)_{\perp}=
D_{\sigma}\partial_{\sigma}x-
{(p,D_{\sigma}\partial_{\sigma}x)_A\over (p,p)_A}p_A-
{(p,D_{\sigma}\partial_{\sigma}x)_S\over (p,p)_S}p_S$. 
The one-form
$(dx, (D_{\sigma}\partial_{\sigma}x)_{\perp})$
is an element of the cotangent space $T^*_{(p,x)}M$ to the
phase space at the point $(p,x)$. It is horizontal in the
sense that its value on ${\partial\over\partial E_A}$,
${\partial\over\partial E_S}$, ${\partial\over\partial\phi_A}$
and ${\partial\over\partial\phi_S}$ is zero. This means
that it is a pullback of some form $\alpha(p,x)$ on the tangent
space to $G$ at the point $\boldpi(p,x)$:
\begin{equation}\label{DifferentialOfH}
 (dx, (D_{\sigma}\partial_{\sigma}x)_{\perp})
=\boldpi^*\alpha(p,x)
\end{equation}
To avoid confusion, we want to stress that this 
form $\alpha(p,x)\in T^*_{\pi(p,x)}G$ 
depends on $(p,x)$ and not just on the projection 
$\boldpi(p,x)$.
That is why we did not use the boldface for $\alpha$.
Given the Eq. (\ref{DifferentialOfH}) for $dH$ and the
symplectic form (\ref{OmegaOnLM}) on $LM$ we can write down
the Hamiltonian vector field:
\begin{eqnarray}\label{HamiltonianVF}
&&\omega^{-1}d(H+\epsilon^2\Delta H)=
{\partial\over\partial\phi_A}+{\partial\over\partial\phi_S}+
\nonumber\\[5pt]&&+
\epsilon^2 \left[(p,D_{\sigma}\partial_{\sigma}x)_A
{\partial\over\partial E_A}+
(p,D_{\sigma}\partial_{\sigma}x)_S
{\partial\over\partial E_S}-
\boldpi^{-1}{\bf\Omega}^{-1}\alpha(p,x)\right]
\end{eqnarray}
\underline{Long term evolution.}
The coefficients of $\partial\over\partial E_A$
and  $\partial\over\partial E_S$ 
describe the evolution of the frequency:
\begin{eqnarray}
E_A(\tau)=E_A(0)+\epsilon^2\int_0^{\tau} d\tau' 
(p(\sigma,\tau'),D_{\sigma}\partial_{\sigma}
x(\sigma,\tau'))_A\nonumber\\[5pt]
E_S(\tau)=E_S(0)+\epsilon^2\int_0^{\tau} d\tau' 
(p(\sigma,\tau'),D_{\sigma}\partial_{\sigma}
x(\sigma,\tau'))_S\nonumber
\end{eqnarray}
We want to study the evolution over the period up 
to the order $\epsilon^2$
therefore we can replace on the right hand side 
$x(\sigma,\tau')$ and $p(\sigma,\tau')$ 
with the unperturbed motion $x_0(\sigma,\tau')$ and 
$p_0(\sigma,\tau')$.

We can now see that $E_A(\tau)$ 
and $E_B(\tau)$ oscillates around $E_A(0)$ and $E_B(0)$. 
Indeed, taking into account the initial condition 
(\ref{SpecialChoiceOfTilt}) we have:
\begin{equation}
\int d\tau' 
(\partial_{\tau'}x(\sigma,\tau'),D_{\sigma}\partial_{\sigma}
x(\sigma,\tau'))_A=
-{1\over 2}\int d\tau' {\partial\over\partial\tau'}
(\partial_{\sigma}x(\sigma,\tau'),\partial_{\sigma}x(\sigma,\tau'))_A
=0
\end{equation}
because of the periodicity. 
Therefore the variations of the frequency do not accumulate over
 time. The initial conditions (\ref{SpecialChoice})
imply that $E_A(0)=1-\epsilon^2
(\partial_{\sigma}x,\partial_{\sigma}x)_A\;+\!$ 
(terms of the higher order in $\epsilon^2$).

But the variation of the shape of the contour does accumulate. 
For $\tau$ of the order $1\over \epsilon^2$
the change in the shape of the contour will be of the order
one. Indeed (\ref{HamiltonianVF}) implies that the projection
of the trajectory on $G$ satisfies:
\begin{equation}\label{ProjectionToBase}
\partial_{\tau}\boldpi(p,x)=
-\epsilon^2{\bf\Omega}^{-1}\alpha(p,x)
\end{equation}
The variation of the geodesic over one period is therefore:
\begin{equation}
\delta \boldpi = -{\bf\Omega}^{-1}\int_0^{2\pi}d\psi\; 
\alpha(p_0,x_0(\psi))
\end{equation}
Again, we neglected the higher order terms in $\epsilon^2$
and replaced all the $(p(\tau),x(\tau))$ on the right hand side 
of (\ref{ProjectionToBase}) by the
unperturbed $p_0(\tau),x_0(\tau)$. Also, following the
notations in (\ref{SymplecticFormOnJacobi}) we replaced the time $\tau$
with the angle $\psi$ parametrizing the geodesic.
Notice that $\int_0^{2\pi}d\psi\;\alpha(p_0,x_0(\psi))$
is the differential of the function on the base $G$ which is
obtained by the integration of $\Delta H$ over $\psi$:
\begin{eqnarray}
&&\int_0^{2\pi}d\psi\;\alpha=
d\;{\bf \overline{\Delta H}},\\[5pt]
&&{\bf\overline{\Delta H}}={1\over 2}\int_0^{2\pi}d\psi
\int _0^{2\pi} d\sigma\;(\partial_{\sigma}x_0(\psi,\sigma),
\partial_{\sigma}x_0(\psi,\sigma))\label{AverageHamiltonian}
\end{eqnarray}
Let us prove it. We have
\begin{equation}
\int d\psi\;\alpha=
\int d\psi d\sigma\; (D_{\sigma}dx_{\perp}(\psi,\sigma),
\partial_{\sigma}x(\psi,\sigma))
\end{equation}
By definition 
$dx_{\perp}=dx-(dx,\partial_{\psi}x)_A\partial_{\psi}x_A+
(dx,\partial_{\psi}x)_S\partial_{\psi}x_S$.
(Remember that in our notations the metric on $S^5$ is negative
definite.)
Therefore:
\begin{eqnarray}
&&\int d\psi \;\alpha=\int d\psi d\sigma\; (D_{\sigma}dx,
\partial_{\sigma}x)-\nonumber\\[5pt]
&&-\int d\psi d\sigma \;
(D_{\sigma}((\partial_{\psi}x,dx)_A\partial_{\psi}x),
\partial_{\sigma}x)_A
-\int d\psi d\sigma \;
(D_{\sigma}((\partial_{\psi}x,dx)_S\partial_{\psi}x),
\partial_{\sigma}x)_S\nonumber
\end{eqnarray}
But the second and the third terms on the right hand side are
zero on the initial conditions (\ref{SpecialChoiceOfTilt}).
Therefore $\int d\psi\;\alpha=d\;{\bf \overline{\Delta H}}$
as we wanted.

Now we can compute the variation of $\boldpi(p,x)$ over the period:
\begin{equation}
 \delta \boldpi= -\epsilon^2 {\bf\Omega}^{-1}d\;
{\bf \overline{\Delta H}}({\boldpi})
\end{equation}
Introducing ${\bf t}=\epsilon^2\tau$ we 
obtain the equation for the secular evolution:
\begin{equation}\label{SecularOnG}
{\partial \boldpi\over\partial {\bf t}}=-{\bf\Omega}^{-1} 
d\;{\bf\overline{\Delta H}}({\boldpi}) 
\end{equation}
In this equation all the letters (except for $d$ and $\partial$)
are boldface, as we wanted. It describes the evolution
of the contour in the moduli space of 
null-geodesics on $AdS_m\times S^n$.

\subsection{Summary.}
{The effective Hamiltonian} is a functional 
on the space of parametrized null-surfaces:
\begin{equation}\label{EffectiveHamiltonian}
\overline{\bf \Delta H}={1\over 2}
\int_0^{2\pi}d\psi \int_0^{2\pi} d\sigma\;
(\partial_{\sigma}x,\partial_{\sigma}x)
\end{equation}
Here $\psi$ is the affine parameter
on the light rays and that the periodicity of
the light rays is $\Delta\psi=2\pi$.
 The remaining
coordinate freedom is in the choice of the
closed contour $\psi=\mbox{const}$, but 
the integral (\ref{EffectiveHamiltonian})
does not depend on this choice. Therefore it is
a functional on the space of parametrized null surfaces. 

{The symplectic form} on the space of
parametrized null-surfaces is
\begin{equation}\label{SFNS}
	{\bf \Omega}=\int d\sigma (dx\wedge D_{\psi} dx)
\end{equation}
This symplectic form has a straightforward geometrical interpretation.
Notice that the space of classical string worldsheets has
a natural symplectic form which is defined
in the following way. 
The deformations of the string worldsheet are described by the
 vector fields $\xi(\sigma,\tau)$. 
The value of the symplectic form on two
infinitesimal deformations $\xi_1$ and $\xi_2$ is
\begin{equation}\label{StringSymplecticForm}
	\Omega_{string}(\xi_1,\xi_2)=
	{\sqrt{\lambda}\over 2\pi} \oint ((\xi_1,*D\xi_2)-
	(\xi_2,*D\xi_1))
\end{equation}
Here $D$ is the covariant differential on the worldsheet, 
the metric on the worldsheet is induced from the spacetime, 
the integral
is taken over a closed spacial contour  
 and the fields $\xi_1$ and $\xi_2$ are chosen  to
preserve the conformal structure on the worldsheet 
(they are originally
defined only up to the vector tangent to the worldsheet.)   
The symplectic form (\ref{SFNS}) on the space
of null-surfaces is the ultrarelativistic limit
of the symplectic form (\ref{StringSymplecticForm}) on 
the phase space of the classical string.
Indeed, when $\epsilon\to 0$ (\ref{StringSymplecticForm}) becomes
\begin{equation}
	\Omega_{string}=
	{\sqrt{\lambda}\over 2\pi\epsilon}\int (dx\wedge D_{\psi}dx)
\end{equation}
As we will explain in Section \ref{FTInterpretation},
this equation justifies our 
definition of the small parameter $\epsilon$ and 
the parametrization $\sigma$. Indeed, the right hand side
agrees on the field theory side with the symplectic structure 
of the continuous limit of the spin chain. The parameter $\sigma$ 
should be 
identified with the number of the site divided by the length
of the chain. 

In the end of this section we will  derive this evolution
equation (\ref{SecularOnG}) directly 
from the inhomogeneous Jacobi equation.
But first we want to rewrite (\ref{SecularOnG}) in a more explicit
form and discuss its interpretation in the dual gauge theory.

\subsection{Explicit evolution equations.}
Here we will realize the moduli space of geodesics
as a quadric in the complex projective space
and write the evolution 
equation (\ref{SecularOnG}) in the explicit form. 
Let us start with the $S^n$ part.
Geodesics on $S^n$ are equators:
\begin{equation}\label{Equator}
x_0(\tau,\sigma)=e_1(\sigma)\cos\tau+ e_2(\sigma)\sin\tau
\end{equation}
They are parametrized by a pair of orthogonal vectors
$e_1$ and $e_2$ modulo the orthogonal transformations
mixing $e_1$ and $e_2$. As a manifold it is
the Grassmanian of two-dimensional planes in
the $n+1$-dimensional space, $G=Gr(2,n+1)$.
Let us introduce  a complex vector
$Z=e_1+ie_2$ in ${\bf C}^{n+1}$. 
It has the properties $(Z,Z)=0$ and
$(\overline{Z},Z)=2$. Given the equator, $Z$ is determined
up to a phase $Z\to e^{i\alpha}Z$. 
Therefore the moduli space of geodesics is a quadric 
in the complex projective space ${\bf CP}^n$ 
given in the homogeneous coordinates $[Z_1:\ldots :Z_{n+1}]$
by the equation $(Z,Z)=0$. 
Similarly, the moduli space of geodesics on $AdS_m$ is
a quadric in ${\bf CP}^m$ given in the homogeneous
coordinates $[Y_{-1},Y_0,\ldots, Y_{m-1}]$ by the equation
$(Y,Y)=Y_{-1}^2+Y_0^2-Y_1^2-\ldots-Y_{m-1}^2=0$.  

In our application we need actually not just the geodesic,
but also the position of the point on it. Therefore we have
to keep the phases of $Z$ and $Y$.
The position of the point of the string in 
$AdS_m\times S^n$  is given by 
$$(x_A,x_S)=(\mbox{Re}\;Y,\mbox{Re}\;Z)$$
and the velocity is 
$$(p_A,p_S)=
(\sqrt{E_A}\;\mbox{Im}\;Y,\sqrt{-E_S}\;\mbox{Im}\;Z)$$
The averaged perturbation Hamiltonian is
\begin{equation}
\overline{\bf \Delta H}={1\over 4}\int d\sigma
\left[(\partial_{\sigma}\overline{Y},\partial_{\sigma}Y)
-(\partial_{\sigma}\overline{Z},\partial_{\sigma}Z)\right]
\end{equation}
with the constraint
\begin{equation}
(\overline{Y},\partial_{\sigma}Y)-
(\overline{Z},\partial_{\sigma}Z)=0
\end{equation}
The symplectic form 
\begin{equation}\label{OmegaYZ}
\Omega={1\over 2i}\int d\sigma ((d\overline{Y}\wedge dY)-
(d\overline{Z}\wedge dZ))
\end{equation}
The Hamiltonian flow (\ref{HamiltonianVF}) 
{\it averaged over the period } is:
\begin{eqnarray}
&\hspace{-30pt}&\partial_{\tau}Y=i\left[\left(1-
{1\over 2}\epsilon^2
(\partial_{\sigma}\overline{Y},\partial_{\sigma}Y)
\right)Y-
{1\over 2}\epsilon^2\partial^2_{\sigma}Y
-{1\over 4}\epsilon^2
(\partial_{\sigma}Y,\partial_{\sigma}Y)\overline{Y}\right]
\nonumber\\[5pt]
&\hspace{-30pt}&\partial_{\tau}Z=i\left[\left(1-
{1\over 2}\epsilon^2
(\partial_{\sigma}\overline{Z},\partial_{\sigma}Z)
\right)Z-
{1\over 2}\epsilon^2\partial^2_{\sigma}Z
-{1\over 4}\epsilon^2
(\partial_{\sigma}Z,\partial_{\sigma}Z)\overline{Z}\right]
\label{SecularOnQuadric}
\end{eqnarray}
The terms proportional to $Y$ and $Z$ are fixed from 
the initial condition (\ref{SpecialChoice}), 
and the terms proportional to $\overline{Y}$ and $\overline{Z}$
are such that $(\partial_{\tau}Y,Y)=0$ and
$(\partial_{\tau}Z,Z)=0$.

\subsection{Interpretation 
in the dual field theory.}\label{FTInterpretation}
To interpret these equations on the field theory
side we have to consider the single trace operators
with large R-charge. In the ``continuum limit''
$Z$ corresponds to the local density of the R charge. 
The operators corresponding to the speeding strings
are ``locally half-BPS'' \cite{MateosMateosTownsend}. 
Therefore the density of the R charge should be 
a decomposable element of $so(6)$ which means that $(Z,Z)=0$.
Following the idea of \cite{Kruczenski} we can interpret
$Z$  as parametrizing
a point on the coadjoint orbit 
of $so(6)$ consisting of the decomposable elements. 
Decomposable elements are those
antisymmetric matrices which can be represented
as an antisymmetric product of two orthogonal vectors 
$e_1\wedge e_2$; then $Z=e_1+ie_2$. 
This orbit
corresponds in the sense of \cite{Perelomov}
to the vector representation of $so(6)$
which lives on the sites of the spin chain. 

Let us now add the AdS part. Consider the orbit of $so(2,4)$
consisting of the elements of the form $Y\wedge \overline{Y}$
where $Y=\tilde{e}_1+i\tilde{e}_2$ with $(Y,Y)=0$
and $(\overline{Y},Y)=2$. Just as a geodesic in  $S^5$ is
defined by $Z$ modulo a phase, a geodesic in
$AdS_5$ is defined by $Y$ modulo a phase.
Roughly speaking, a pair of functions 
$(Z(\sigma),Y(\sigma))$ where both $Z(\sigma)$ and $Y(\sigma)$ 
are defined modulo local phase rotations 
(independent for $Z$ and $Y$) define a null-surface 
in $AdS_5\times S^5$. But there is a subtlety. 
For the corresponding surface to be null we have to be able
to fix the relative phase of $Y$ and $Z$ in such a way
that
\begin{equation}\label{RelativePhase}
(\overline{Z},\partial_{\sigma}Z)=
(\overline{Y},\partial_{\sigma}Y)
\end{equation}
This imposes the following integrality condition on the functions
$Y(\sigma)$ and $Z(\sigma)$. 
Let us consider a two-dimensional surface $D_Z$ in
${\bf CP}^6$ such that its boundary is the contour
$[Z(\sigma)]$ and a two-dimensional surface
$D_Y$ in ${\bf CP}^{2+4}$ such that its boundary
is the contour $[Y(\sigma)]$. The integrality condition
is that the symplectic area of $D_Y$ should be equal
to the symplectic area of $D_Z$ plus an integer.
On the field theory side this integrality condition
corresponds to the cyclic invariance of the trace.

To summarize, let us consider two functions 
$[Y]:S^1\to {\bf CP}^{2+4}$ and $[Z]:S^1\to {\bf CP}^6$
satisfying $(Y,Y)=(Z,Z)=0$ and 
the integrality condition described above.
 The integrality condition guarantees that we can
lift $[Z]$ and $[Y]$ to the functions
$Y:S^1\to {\bf C}^{2+4}$ and $Z:S^1\to {\bf C}^6$
satisfying (\ref{RelativePhase}). 
Let us fix such a lift modulo an overall phase
$(Y,Z)\sim e^{i\phi(\sigma)}(Y,Z)$.
This data
determines the null surface in $AdS_5\times S^5$
 corresponding to the
Yang-Mills operator with the anomalous dimension
\begin{equation}\label{AnomalousDimensionInYZ}
\epsilon{\sqrt{\lambda}\over 8\pi}\int d\sigma \left(
(\partial_{\sigma}\overline{Z},\partial_{\sigma}Z)-
(\partial_{\sigma}\overline{Y},\partial_{\sigma}Y)\right)
\end{equation}
In this formula we have restored the
coefficient  ${\sqrt{\lambda}\over 4\pi\epsilon}$ from
Eq. (\ref{WorldsheetAction}).
The integral does not depend on the ``overall'' phase 
of $(Y,Z)$.  

 The precise relation  between $\epsilon$ and 
$\lambda$ can be obtained by computing the conserved charges. 
Consider a Killing vector field $U$ on $S^5$. 
We have
\begin{equation}
\delta_U x^i=u^{ij}x^j
\end{equation}
where  
$x^i$, $i=1,\ldots,6$ denote a unit vector 
representing the point of $S^5$ and $u^{ij}$
is an antisymmetric matrix corresponding to the
symmetry $U$. 
Let us compute the corresponding conserved charge
to the first order in $\epsilon$. We have:
\begin{equation}
Q_U={1\over\epsilon}{\sqrt{\lambda}\over 2\pi}
\int_0^{2\pi} d\sigma\; u^{ij}x_0^j(\tau,\sigma)
\partial_{\tau} x_0^i(\tau,\sigma)
\end{equation}
By definition $x_0(\tau,\sigma)$ should belong
to the geodesic specified by $Z(\sigma)$, and
$\partial_{\tau}x^i=
\left({i\over 2}Z\wedge\overline{Z}\right)^{ij}x^j$.
This means that the charge is:
\begin{equation}
Q_U={1\over\epsilon}{\sqrt{\lambda}\over 2\pi}
\int_0^{2\pi} d\sigma\; 
\left(u,{i\over 2}\overline{Z}\wedge Z\right)
\end{equation}
But ${i\over 2} \overline{Z}\wedge Z$ 
should
be the local density of the R charge. Therefore
we  identify
\begin{equation}
\epsilon={\sqrt{\lambda}\over 2\pi (L/2\pi)}
\end{equation}
where $L$ is the length of the spin chain (the number 
of operators under the trace.) Substitution of $\epsilon$
in (\ref{AnomalousDimensionInYZ}) gives:
\begin{equation}
\Delta={1\over 16\pi^2}{\lambda\over (L/2\pi)}
\int_0^{2\pi} d\sigma \left(
(\partial_{\sigma}\overline{Z},\partial_{\sigma}Z)-
(\partial_{\sigma}\overline{Y},\partial_{\sigma}Y)\right)
\end{equation}
This is a functional on the space
of contours $(Y(\sigma),Z(\sigma))$ in ${\bf C}^{12}$,
subject to the constraints $|Y|^2=|Z|^2=2$ and
$(\overline{Z},\partial_{\sigma}Z)=
(\overline{Y},\partial_{\sigma}Y)$ and defined
up to an overall phase 
$(Y(\sigma),Z(\sigma))\to e^{i\phi(\sigma)}
(Y(\sigma),Z(\sigma))$. The symplectic structure on this
space is given in (\ref{OmegaYZ}).

\subsection{Comment on the special case when $\Sigma(0)$ is
 generated by the orbits of $V$.}
In the special case when $\Sigma(0)$ is
 generated by the orbits of $V$  the 
anomalous dimension can be computed in two different ways.
One way is to compute the conserved charge
corresponding to $V$ as was done in 
\cite{FT02}. The other way suggested in \cite{Kruczenski} 
is to study the secular evolution of $\Sigma(\epsilon)$ and
find the Hamiltonian governing this evolution.
The two methods give the same result for the
following reason. 
The constraint
 $(\partial_{\tau}x)^2+\epsilon^2(\partial_{\sigma}x)^2=0$
says that the total perturbed Hamiltonian $H_0+\epsilon^2\Delta H$
should be zero. The ``effective'' Hamiltonian governing
the secular drift is obtained by the averaging of
$\Delta H$ over the period. Because of the constraint
we have $\epsilon^2 \Delta H=-H_0$. 
But in the vicinity of $\Sigma(0)$ we have $H_0$ equal
to the charge $Q_V$ up to the terms 
of the higher order in the deviation from $\Sigma(0)$. 
(This follows from the fact that the Hamiltonian
flow of $H_0$ on $\Sigma(0)$ is the translation by $V$.)

\subsection{Direct derivation from the Jacobi equation.}
We derived (\ref{SecularOnG}) and (\ref{SecularOnQuadric})
using the Hamiltonian formalism. Here we will give
a direct derivation from the inhomogeneous Jacobi equation.

Let us study the inhomogeneous 
Jacobi equation in the special
case of AdS times a sphere:
\begin{equation}\label{InhomogeneousJacobiA}
D_{\tau}^2\eta-R(\partial_{\tau}x,\eta)\partial_{\tau}x=
D_{\sigma}\partial_{\sigma}x
\end{equation}
We can decompose $\partial_{\tau}x$ 
as a sum of the vector $\partial_{\tau}x_{AdS_5}$ in the tangent
space to $AdS_5$ and the vector $\partial_{\tau}x_{S^5}$ 
in the tangent space to $S^5$, 
$\partial_{\tau}x=\partial_{\tau}x_{AdS_5}+
\partial_{\tau}x_{S^5}$. 
We denote 
$\widetilde{\partial_{\tau}x}=
\partial_{\tau}x_{AdS_5}-\partial_{\tau}x_{S^5}$. 
We will need the following
representation for $D_{\sigma}\partial_{\sigma}x$:
\begin{equation}\label{DdxLemma}
D_{\sigma}\partial_{\sigma}x=D_{\tau}\xi+
\alpha(\sigma,\tau) \partial_{\tau}x+
\beta(\sigma,\tau)\widetilde{\partial_{\tau}x}
\end{equation}
where $\xi$ is a Jacobi field 
orthogonal to both $\partial_{\tau}x$ and 
$\widetilde{\partial_{\tau}x}$ 
and $\alpha(\tau)$ and $\beta(\tau)$
are some functions. Indeed, let us consider the projection
of the geodesic on $S^5$. The geodesic on $S^5$ is an equator: 
\begin{equation}
x(\tau,\sigma)=e_1(\sigma)\cos\tau+ e_2(\sigma)\sin\tau
\end{equation}
where $(e_1(\sigma),e_1(\sigma))=(e_2(\sigma),e_2(\sigma))=1$
and $(e_1(\sigma),e_2(\sigma))=0$. We have
\begin{equation}
D_{\sigma}\partial_{\sigma}x=
(e_1''(\sigma)\cos\tau
+e_2''(\sigma)\sin\tau)_{||}
\end{equation}
where the index $||$ means that we have to project 
to the tangent space of $S^5$ along the radial direction. 
It is enough to consider
this equation at $\sigma=0$. Let us decompose the second
derivative of $e_i$, $i=1,2$ in the components $a_{i,tang}$ and
$a_{i,norm}$ parallel to the
plane $(e_1,e_2)$ and the components 
$(e_i'')_{vert}$ 
perpendicular to this plane:
\begin{eqnarray}
e''_1= a_{1,t} e_2 + a_{1,n} e_1
+(e_1'')_{vert}  
\\[5pt]
 e''_2= a_{2,t} e_1 + a_{2,n} e_2
+(e_2'')_{vert}
\end{eqnarray}
The second covariant derivative is:
\begin{eqnarray}
&&D_{\sigma}\partial_{\sigma} x(\tau,\sigma)=\nonumber
\\[5pt]&&=
\left(a_{1,t}\cos^2\tau-a_{2,t}\sin^2\tau+
(a_{2,n}-a_{1,n})\cos\tau\sin\tau\right)
\partial_{\tau}(e_1\cos\tau+e_2\sin\tau)
+\nonumber
\\[5pt]&&+
(e_1'')_{vert}\cos\tau+(e_2'')_{vert}\sin\tau
\end{eqnarray}
The analogous expression holds for the $AdS_5$-component of
$D_{\sigma}\partial_{\sigma}x$. 
But $(e_1'')_{vert}\cos\tau+(e_2'')_{vert}\sin\tau=
\partial_{\tau}
((e_1'')_{vert}\sin\tau-(e_2'')_{vert}\cos\tau)$ and
$$\xi=(e_1'')_{vert}\sin\tau-(e_2'')_{vert}\cos\tau$$ is a Jacobi
field. This proves (\ref{DdxLemma}). Notice that $\xi$ and 
$D_{\tau}\xi$ 
are orthogonal to both $\partial_{\tau}x$ and 
$\widetilde{\partial_{\tau}x}$.
We can now present a solution to the equation 
(\ref{InhomogeneousJacobiA}):
\begin{equation}
\eta={1\over 2}\tau \xi + A\partial_{\tau}x+
B\widetilde{\partial_{\tau}x}
\end{equation}
where $A$ and $B$ satisfy ${\partial^2 A\over \partial \tau^2}=\alpha$
and ${\partial^2 B\over\partial \tau^2}=\beta$. It is important
that both $A$ and $B$ can be chosen periodic functions of $\tau$. 
This is true for $B$:
\begin{equation}
\int d\tau \beta= \int d\tau  \;
(\partial_{\tau}x,D_{\sigma}\partial_{\sigma}x)=
-{1\over 2}\int d\tau\; 
\partial_{\tau}(\partial_{\sigma}x,\partial_{\sigma}x) =0
\end{equation}
and also for $A$, because 
\begin{eqnarray}
&&\int d\tau \; \alpha = 
\int d\tau  \; (\widetilde{\partial_{\tau}x},
D_{\sigma}\partial_{\sigma}x)=\\[5pt]&&=
-{1\over 2}\int d\tau\; 
\partial_{\tau}\left[
(\partial_{\sigma}x,\partial_{\sigma}x)_{AdS_5}-
(\partial_{\sigma}x,\partial_{\sigma}x)_{S^5}\right]=0
\end{eqnarray}
since the projections of $x$ to $AdS_5$ and $S^5$ are
both periodic.
Therefore we see that $\eta$ can be chosen as
a sum of the linearly growing term and the oscillating terms.
The linearly growing term is ${1\over 2}t\xi$ where
$\xi$ is a Jacobi field satisfying 
$D_{\tau}\xi=D_{\sigma}\partial_{\sigma}x$.
This linear term is responsible for the secular evolution.

\section*{Acknowledgments}
I would like to thank S.~Moriyama for discussions
and M.~Kruczenski and A.~Tseytlin for the correspondence.
This research was supported by the Sherman Fairchild 
Fellowship and in part
by the RFBR Grant No.  03-02-17373 and in part by the 
Russian Grant for the support of the scientific schools
No. 00-15-96557.

\end{document}